\documentclass[reprint,amsmath,amssymb,aps,prl,]{revtex4-1}
\usepackage[utf8]{inputenc}
\usepackage{graphicx}% Include figure files
\usepackage{dcolumn}% Align table columns on decimal point
\usepackage{bm}% bold math
\usepackage{amsmath}
\usepackage{amssymb}
\usepackage[justification=raggedright]{caption}
\usepackage{verbatim}
\usepackage{float}
\begin{document}

%\preprint{}

\title{Reflected attosecond pulse radiation from moving electron layers}


\author{Mykyta Cherednychek and Alexander Pukhov \\ \normalsize{\textsl{Institut f\"ur theoretische Physik, Heinrich-Heine-Universit\"at D\"usseldorf}}}
%\thanks{}
\affiliation{}


\date{\today}

\begin{abstract}
	With the generation of high order harmonics (HHG) on the plasma surface it is possible to turn the laser pulse into a train of attosecond or even zeptosecond pulses in the back radiation.
	These attosecond pulses may have amplitude several orders of magnitude larger than that of the laser pulse under appropriate conditions. We study this process in detail, 
	especially the nanobunching of the plasma electron density. We derive the analytical expression that describes the electron density profile and obtain a good agreement with 
	particle-in-cell simulations. We investigate the most efficient case of HHG at moderate laser intensity ($a_0=10$) on the over dense plasma slab with an exponential profile per-plasma.
	Subsequently we calculate the spectra of single attosecond pulses from back radiation
	using our expression for density shape in combination with the equation for spectrum of  nanobunch radiation. 
\end{abstract}

%\keywords{}

\maketitle

\section{Introduction} 
	The development of laser technology showed an immense progress in last decade \cite{Yanovsky, MT11, SSJ, MMKS}. 
	This progress offers an opportunity to study the new physics phenomena of laser plasma interaction. 
	One of the most important processes in this field is the generation of high order harmonics (HHG), which is studied very intensely today.
	As the minimum achievable duration of  laser pulses was reducing with time, 
	it is interesting whether the generation even shorter pulses (in attosecond or even zeptosecond range) is possible.
	The reduction of the pulse duration and the radiation wavelength would open new potential applications. This is the motivation of the investigation of HHG. 
	The most efficient method of HHG is the  interaction process of the high contrast laser pulses \cite{TQG} with solid density targets. 
	The pedestal of the pulse ionizes the surface and the main pulse interacts  with overdense plasma electrons, while ions remain nearly immobile during the short pulse duration.  
	One distinguishes two main HHG mechanisms in this case: coherent wake emission (CWE) \cite{TQG, QTM} and “relativistically oscillating mirror” (ROM)  
	\cite{GPSB, BGP, DZG, DKB, HHM}. CWE is caused by fast Brunel electrons \cite{Brunel} which excite the plasma oscillations at the local plasma frequencies. 
	Thus there is no harmonics behind the maximal plasma frequency in the case of CWE. 
	This process dominates  for non-relativistic laser intensities $a_0\lesssim 1$. For  $a_0\gg1$ the harmonics are generated mostly via ROM. 
	In this case the electron layer at the plasma surface acts as a mirror that oscillates at relativistic velocities 
	and generates high order harmonics via Doppler effect moving toward the incident wave. By this process there is no limit of frequency like by CWE, so higher harmonics can be generated.  
	The first theoretical description of ROM claimed that the intensity spectrum envelope of reflected wave can be described  by $I(n)\propto n^{-5/2}$ 
	until “roll over” frequency $\omega_r$ proportional to $4\gamma^2$, where $n$ is the harmonic order and $\gamma$ is the relativistic gamma factor \cite{GPSB}. 
	Later this theory was improved, especially the acceleration of the reflecting layer was taking into account. This results to the power law   $I(n)\propto n^{-8/3}$ and 
	$\omega_r\propto\gamma^3$ \cite{BGP}. 
	This model assumes the existence of so called apparent reflection point (ARP) where the transverse electric field vanishes. This model was experimentally proved \cite{DZG, DKB, HHM}.  
	Most recently  another HHG mechanism was discovered.
	Using p-polarized oblique incident pulse with $a_0\gg1$ one can cause the formation of extremely dense electron nanobunches under appropriate conditions. 
	These bunches irradiate attosecond pulses with intensities much larger comparing to incident pulse \cite{BP2, BP}. 
	That means that the boundary condition assumed in \cite{BGP} corresponding to ARP  fails and thus the ROM  theory can't be applied in this case. 
	This process is called coherent  synchrotron emission  (CSE). 
	The reflected radiation in case of CSE in characterized by the power law $I(n)\propto n^{-4/3}$  or  $I(n)\propto n^{-6/5}$ which is flatter comparing to ROM \cite{BP2, BP}. 
	The corresponding experiments can be found in Ref. \cite{DRY, DCR, YDC}. 
	Detailed numerical investigation of the case of p-polarized oblique incidence in Ref. \cite{GKMS}  
	figures out that the ROM model can be violated when the  similarity parameter $S=n/a_0$
	(where $n$ is the electron density given in units of the critical density $n_c$ 
	and $a_0$ is the dimensionless laser amplitude \cite{GP}) is smaller than five.  
	The authors of \cite{GKMS} present  a new  relativistic electronic spring (RES) model for $S>5$.  
\subsection{PIC simulation of the HHG process}
 	For our simulations we use one dimensional version of PIC code called Virtual Laser Plasma Laboratory \cite{vlpl}. 
	In our geometry the incident wave comes from the left hand side of the simulation box and propagates along $x$-axis. 
	The wave is p-polarized and the electric field component oscillates along the $y$-axis. 
	The plasma is located at the right hand side of simulation box.  
	It is also possible to describe the interactions where oblique  incidence is used with our code. 
	Let $\theta$ be the angle of incidence in laboratory frame and consider some frame moving along $y$-axis with velocity $V=c\sin\theta$. 
	Lorenz transformations verify that in this frame the laser is normally incident (see \cite{LL} for more details). 
	At the same time the whole plasma moves in $y$-direction in this frame. 
	Thus, attributing some initial velocity to plasma in our simulation we are working in the moving frame. 
	If we need the results in laboratory frame, we have to transform the values  obtained from the simulation via   Lorenz transformation. 
	Consequently we obtain results that correspond to the process with oblique incidence. 
	We use the incident field $E_i(t)$ of duration $T=10\lambda/c$, that is given by 
	\begin{align}
		\nonumber
		E_i(t)=&\frac{1}{4}\left(1+\tanh\left(\frac{t}{\Delta t}\right)\right)
		\\
		\nonumber
		&\times\left(1-\tanh\left(\frac{t-T}{\Delta t}\right)\right)\sin(2\pi t),
	\end{align}
	where $\Delta t = \lambda/4$. 
	Further we use the plasma exponential density ramp  for $x<0$. For $x>0$ density remains constant. 
	\begin{align}
		\label{rump}
		n(x)=\left \lbrace 
		\begin{aligned}
			&n_0e^{\frac{x}{\sigma}}\qquad\text{for}\quad x<0\\
			&n_0\qquad\quad\,\text{for}\quad x>0
		\end{aligned}
		\right ..\\
		\nonumber
	\end{align}
	Assuming that the ions are at rest during the whole interaction  process we consider only the interaction between the electrons and the incident wave. 
	In the simple case of normal incidence there are two forces acting on particles along $x$-axis. 
	The electrostatic force proportional to $E_x$ and laser  ponderomotive force oscillating with $2\omega$ (twice of the laser frequency).  
	Thus the plasma surface oscillates with the half of the laser period. 
	In the case of oblique incidence of p-polarized wave there is additional longitudinal component of the electric field  oscillating at frequency $\omega$ and acting on surface.  
	Consequently the interaction becomes even more complicated  which leads to stronger oscillations on plasma surface containing both $\omega$ and $2\omega$ mods.  
	
	As we can see  as soon as the electrons are puled back by the electrostatic force they form a thin nanobunch that reaches velocity close to $c$. 
	In this case the generation of high harmonics is possible.    

\section{Density profile of a thin electron layer}

	In this section we derive two different analytic expressions for two different cases, 
	which roughly describe the electron density profile at the intervals where the sharp spikes appear.  
	The starting point of our calculations is the approximation of  the electron phase space distribution at these intervals. 
	As we will see later this distribution depends on the propagation  velocity $\dot x_0(t)$ of given electron layer.    

	Firstly let us consider the case of slow  $\dot x_0(t)\ll c$ electron bunch.  
	\begin{figure}[htbp]
		\centering
		\includegraphics[width=7cm]{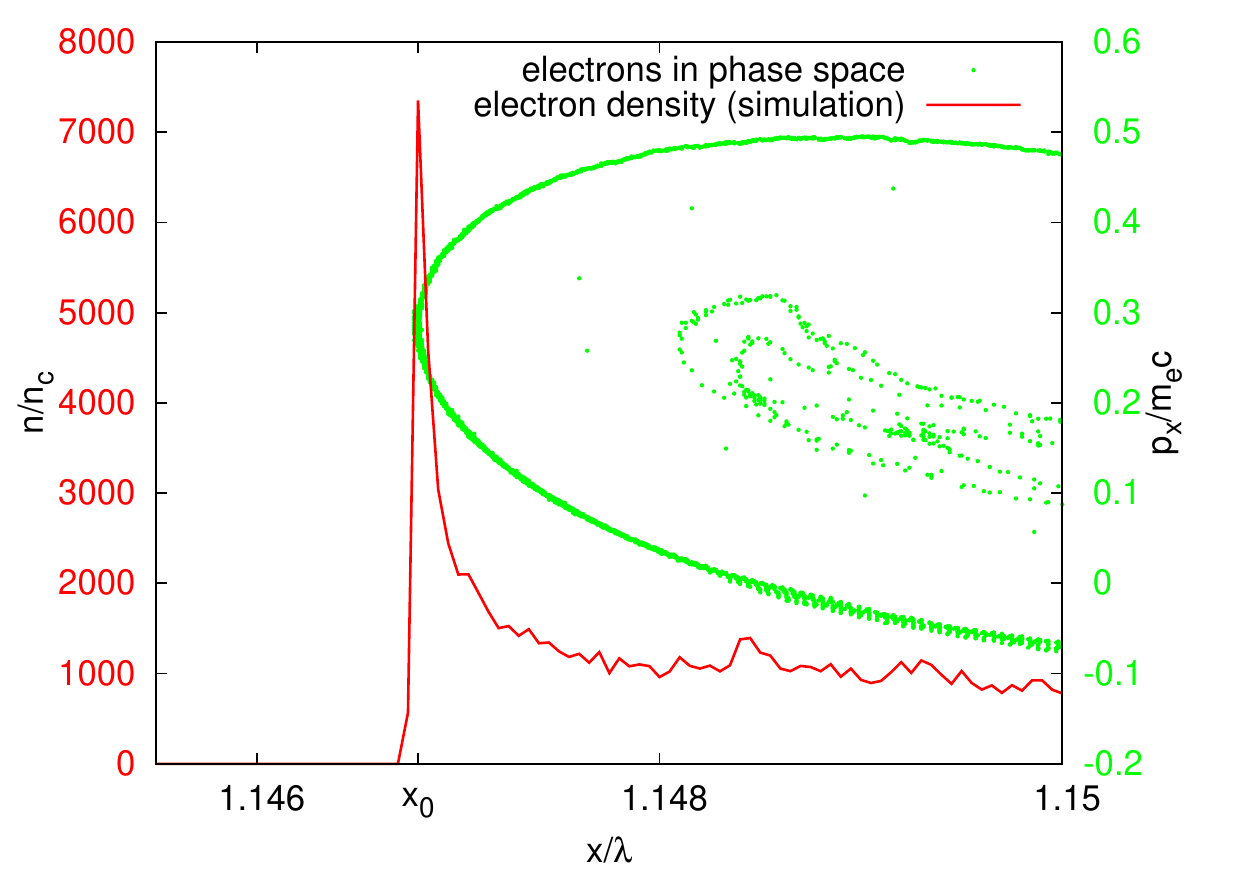}
		\caption{\small Electron density (red) and electrons in $x$-$p_x$-plane (green). $x_0$ is the position of the maximal density. 
		Simulation parameters: initial plasma density $n_0 = 38.9 n_c$; $\sigma=1.1836\cdot10^{-3}\lambda$(Laboratory frame), Pulse with dimensionless amplitude $a_0 = 10$
		and p-polarized oblique incidence at $57^\circ$ angle has the  wave length $\lambda = 820 $nm.}
		\label{dens1_377}
	\end{figure}	
	In Fig. \ref{dens1_377} the electron density and its distribution in $x$-$p_x$-phase space at a certain time  are visualized. 
	Let's claim that the curve in phase space is described by the function $x(p)$ at some small interval close to the density spike. 
	Obviously, $x_0$ is the local minimum of this function that coincides with the position of the spike.  
	In fact, we have always a spike of electron density at the point, 
	where the function $x(p)$ exhibits the local extreme value.  
	The idea that gives us the staring point for our calculations is the following. We can locally describe the given curve in phase space as a parabola:    
	\begin{equation}
		x(p,t)=x_0(t)+\alpha(t)(p-p_0(t))^2.
		\label{x(p)}
	\end{equation}	
	The point ($x_0(t)$,~$p_0(t)$) corresponds to the local minimum. 
	In order to simplify the notation, we drop the time dependence and set $p_0=x_0=0$. Then we have
	\begin{equation*}
		x(p)=\alpha p^2.
	\end{equation*}

\begin{figure}[htb]
		\centering
		\includegraphics[width=7cm]{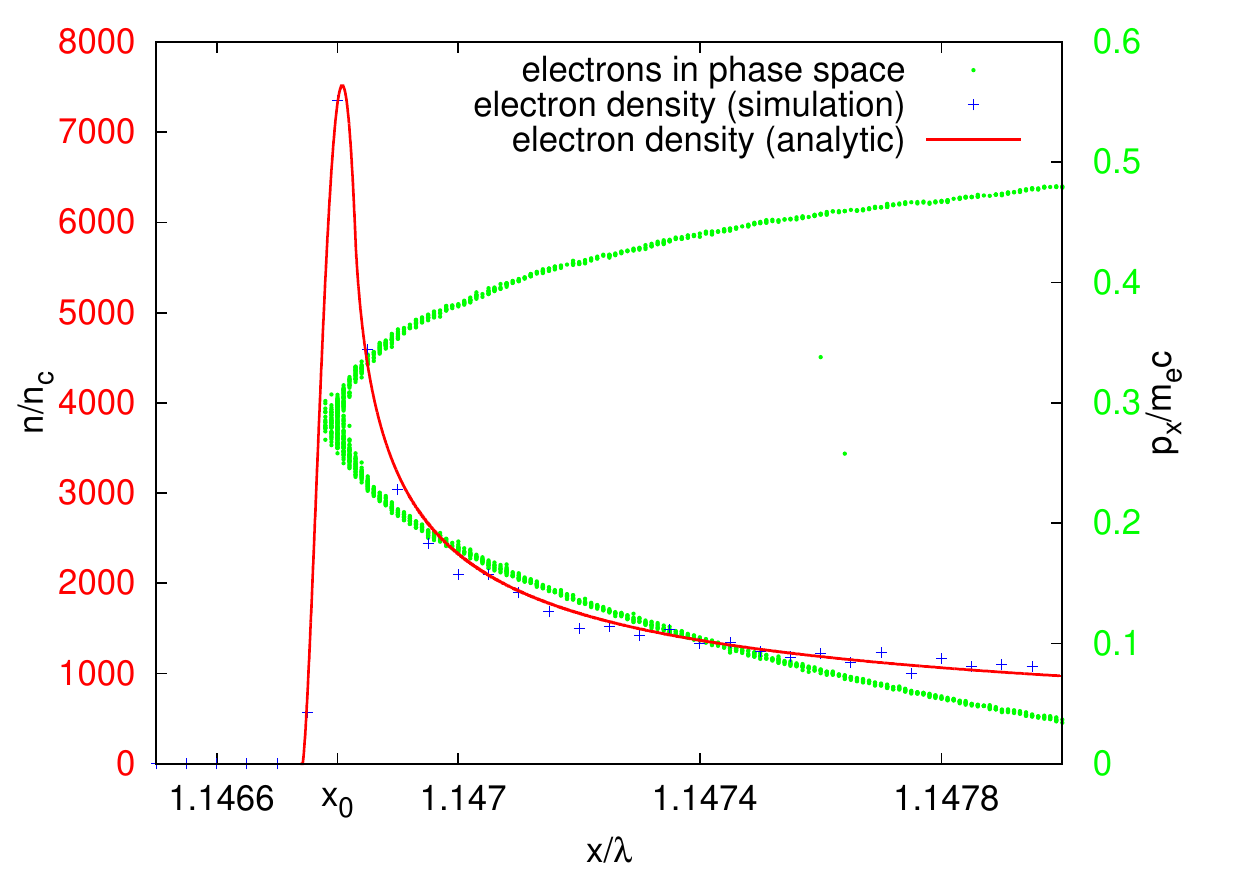}
		\caption{\small Electron density taken from the simulation (blue) and calculated analytically via (\ref{profile_sqrt}) (red),  
		with same simulation parameters compared to Fig. \ref{dens1_377}, $\Delta x=0.0012\lambda$; $a=4.4\cdot10^{-5}$ (simulation frame).}
		\label{plot_prof1}
	\end{figure}

	We consider some short interval $\Delta x$ where this assumption makes sense. 
	The distribution function of the electrons is then given by
	\begin{align}
		f_a(x,p)&={\cal C}\delta_a\left(x-\alpha p^2\right),
		\label{phase_space1}
	\end{align}
	where ${\cal C}$ is a normalization constant and $\delta_a$ is defined as 
	\begin{align}
		\nonumber
		g_a(x)&\equiv\frac{3}{4a}\left(1-\frac{x^2}{a^2}\right)\\
		\nonumber
		\delta_a(x)&\equiv\left \lbrace
		\begin{aligned}
			&g_a(x)\quad\text{for}\quad x\in [-a,a] \\
			& 0\qquad \text{otherwise}
		\end{aligned},
		\right .
	\end{align}	
	with the property 
	\begin{equation}
		\lim_{a\rightarrow0}\delta_a(x)=\delta(x).
		\label{delta_conv}
	\end{equation}	
	The parameter $a$ describes the width of $\delta_a$, which means that $a>0$ is required.
	In order to get the expression of density we have to perform the integration in momentum space
	\begin{equation}
		n_a(x)=\int dpf_a(x,p).
		\label{dens_int}
	\end{equation}
	We have to be careful with integration boundaries since  $\delta_a$ is the bounded support function.  As a result we obtain: 
	\begin{align}
		\nonumber
		n_a(x)=\qquad\qquad\qquad\qquad\qquad\qquad\qquad\qquad\qquad\quad\\
		\nonumber
		\left \lbrace
		\begin{aligned}
			& \frac{2{\cal C}}{5a^3\sqrt{\alpha}}\left(3a^2-2x^2+ax\right)\sqrt{x+a} \quad\text{for}\quad x\in [-a,a]\\
			& \frac{2{\cal C}}{5a^3\sqrt{\alpha}}\biggr(\left(3a^2-2x^2\right)\left(\sqrt{x+a}-\sqrt{x-a}\right)\\
			&\qquad\quad +ax\left(\sqrt{x+a}+\sqrt{x-a}\right)\biggr)~\,\text{for}\quad x>a\\
			& \qquad0\qquad\qquad\qquad\qquad\qquad\qquad~\,\text{for}\quad x<-a.
		\end{aligned}
		\right .\\
		\label{profile_sqrt_C}
	\end{align}
	In order to calculate the constant ${\cal C}$, we first write an equation for the number of particles in the interval $[-a:\Delta x]$ 
	bei integrating the density on this interval
	\begin{equation}
		N_{a,\Delta x}={\cal C}\int_{-a}^{\Delta x}dx~n_a(x)\stackrel{a\ll\Delta x}{=}2{\cal C}\sqrt{\frac{\Delta x}{\alpha}}.
		\label{N}
	\end{equation}
	Further we solve the obtained equation for ${\cal C}$ and insert it into equation (\ref{profile_sqrt_C}). Finally we obtain the expression for electron density profile
	\begin{align}
		\nonumber
		n_{a}(x)=\qquad\qquad\qquad\qquad\qquad\qquad\qquad\qquad\qquad\qquad\\
		\nonumber
		\left \lbrace
		\begin{aligned}
			& \frac{N_{a,\Delta x}}{5a^3\sqrt{\Delta x}}\left(3a^2-2x^2+ax\right)\sqrt{x+a} \quad\text{for}\quad x\in [-a,a]\\
			& \frac{N_{a,\Delta x}}{5a^3\sqrt{\Delta x}}\biggr(\left(3a^2-2x^2\right)\left(\sqrt{x+a}-\sqrt{x-a}\right)\\
			&\qquad\quad +ax\left(\sqrt{x+a}+\sqrt{x-a}\right)\biggr)\quad~\text{for}\quad x>a\\
			& \qquad0\qquad\qquad\qquad\qquad\qquad\qquad\quad~\text{for}\quad x<-a.
		\end{aligned}
		\right .\\
		\label{profile_sqrt}
	\end{align}	
	Note that the parameter $\alpha$ cancels, so it doesn't affect the density profile.
	In Fig. \ref{plot_prof1} we see that the density described with (\ref{profile_sqrt}) agrees very well with simulation results.
	We call the case where $\dot x_0(t)\ll c$ is valid ``parabolic case''.
	We chose quite a small value for $a$ because  we are dealing with a very big and sharp spike in this example. 
	This is the case since we use a strong laser pulse and very small cell size ($5\cdot 10^{-5}\lambda$).

\begin{figure}[htbp]
		\centering
		\includegraphics[width=8.6cm]{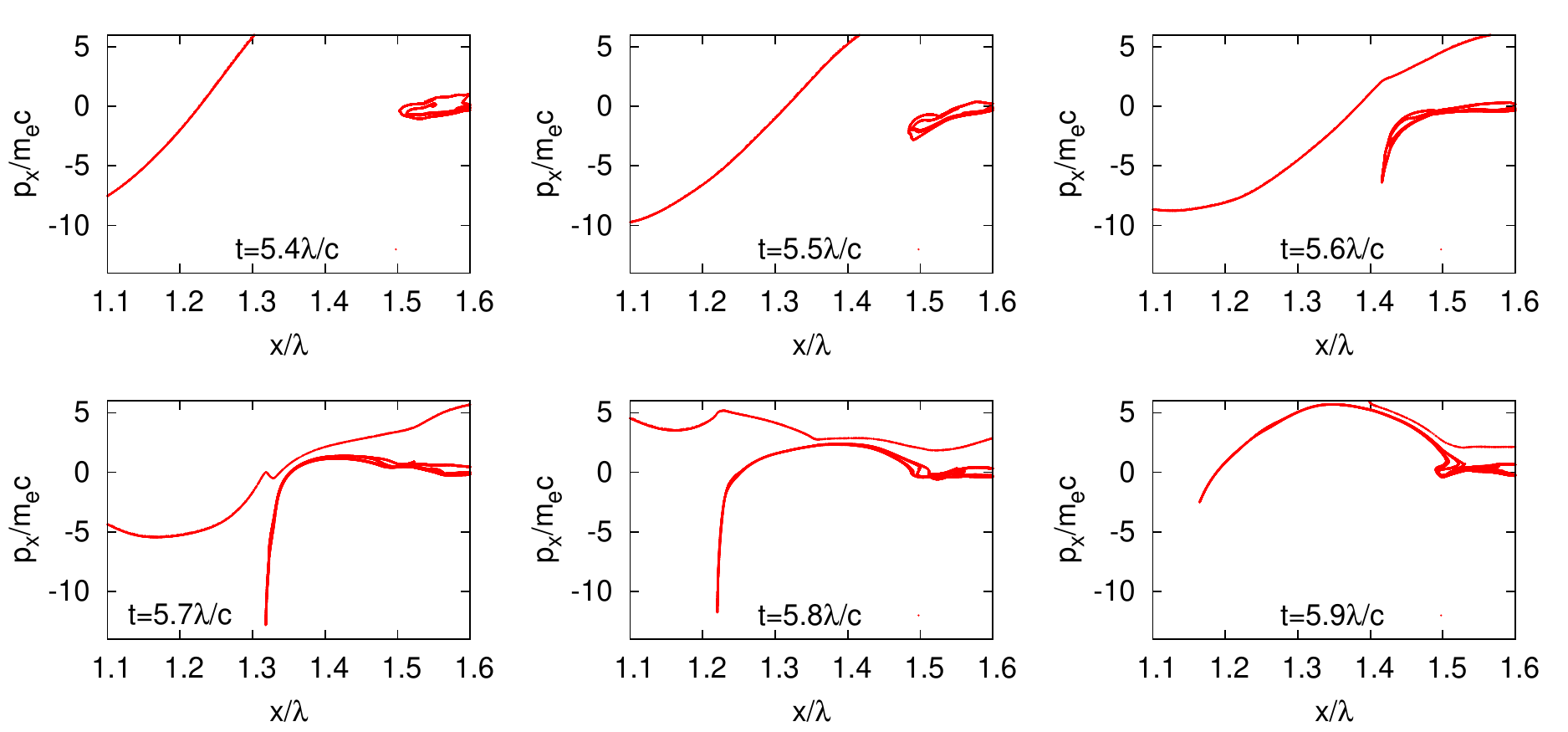}
		\caption{\small Electrons in $x$-$p_x$-plane taken from the simulation to different times $t$ during the process of nanobunching.
		Simulation parameters: initial plasma density $n_0 = 100 n_c$; $\sigma=0.4\lambda$ (laboratory frame), Pulse with dimensionless amplitude $a_0 = 10$
		and p-polarized oblique incidence at $50^\circ$ angle has the  wave length $\lambda = 820 $nm.}
		\label{xpx2}
	\end{figure}

	Now we discuss another case  with $\dot x_0(t)\rightarrow c$. 
	Consider the phase space evolution taken from the other simulation illustrated in Fig. \ref{xpx2}. 
	At the beginning by $t=5.4\lambda/c$ the momentum is close to zero and the distribution is parabolic as expected.  
	Further, as soon as the electron bunch is puled back by the electrostatic force,  
	the negative momentum of the bunch growth constantly with time and the distribution changes  reminding a kind of ``whip" between $t=5.7\lambda/c$ and  $t=5.8\lambda/c$. 
	The extremely dense electron nanobunch reaches the velocity close to $c$ during this period.
	In this case the phase space distribution can be roughly fitted with an exponential function 
	\begin{equation}
		x_p(p, t)=x_0(t)+e^{\alpha(t)\left( p-p_0(t)\right)}.
		\label{exp2}
	\end{equation}			
	As we will show later the incident angle and the density gradient used here are optimal for producing the most intense attosecond pulse. 
	Given phase space distribution belongs to the nanobunch that radiates this pulse.     
	Like in the previous case we drop the time dependence and set $p_0=x_0=0$. Then we have 

	\begin{equation}
		x_p(p)=e^{\alpha p}
		\label{exp}
	\end{equation}
	and the distribution function:
	\begin{align}
		f_a(x,p)&={\cal C}\delta_a\left(x-e^{\alpha p}\right).
		\label{phase_space2}
	\end{align}
	The density can be given by
	\begin{equation}
		n_a(x)=\int_{-p_\text{cut}} dpf_a(x,p)
		\label{dens_int2}
	\end{equation}	
	in this case, since we should take into account that the  momentum of the electrons is limited by some amount $p_\text{cut}$. 
	Further we calculate the number of particles on some interval $[x_\text{min}:x_\text{max}]$, where $x_\text{min}=e^{-\alpha p_\text{cut}}$, in order to obtain  ${\cal C}$.
	Finally we calculate:
	\begin{align}
		\nonumber
		n_a(x)=\qquad\qquad\qquad\qquad\qquad\qquad\qquad\qquad\qquad\qquad\\
		\nonumber
		\left \lbrace
		\begin{aligned}
			\nonumber
			&\frac{3N}{4a^3\ln{\left(\frac{x_\text{max}}{x_\text{min}}\right)}}
			\\
			\nonumber
			&~\times\biggr((x+a)\left(x+(x-a)\left(\frac{1}{2}+\ln\left(\frac{x_\text{min}}{x+a}\right)\right)\right)
			\\
			\nonumber
			&\,+x_\text{min}\left(\frac{1}{2}x_\text{min}-2x\right)\biggr)~\text{for}~ x\in [x_\text{min}-a,x_\text{min}+a]
			\\
			\nonumber
			& \frac{3N}{4a^3\ln{\left(\frac{x_\text{max}}{x_\text{min}}\right)}}
			\\
			\nonumber
			&\,\times\left(2ax-(x^2-a^2)\ln\left(\frac{x+a}{x-a}\right)\right)~\text{for}~x>x_\text{min}+a
			\\
			\nonumber
			& \qquad0\qquad\qquad\qquad\qquad\qquad\qquad\quad\!\text{for}~x<x_\text{min}-a
		\end{aligned}
		\right .\\
		\label{profile_exp}
	\end{align}

\begin{figure}[htb]
		\centering
		\includegraphics[width=7cm]{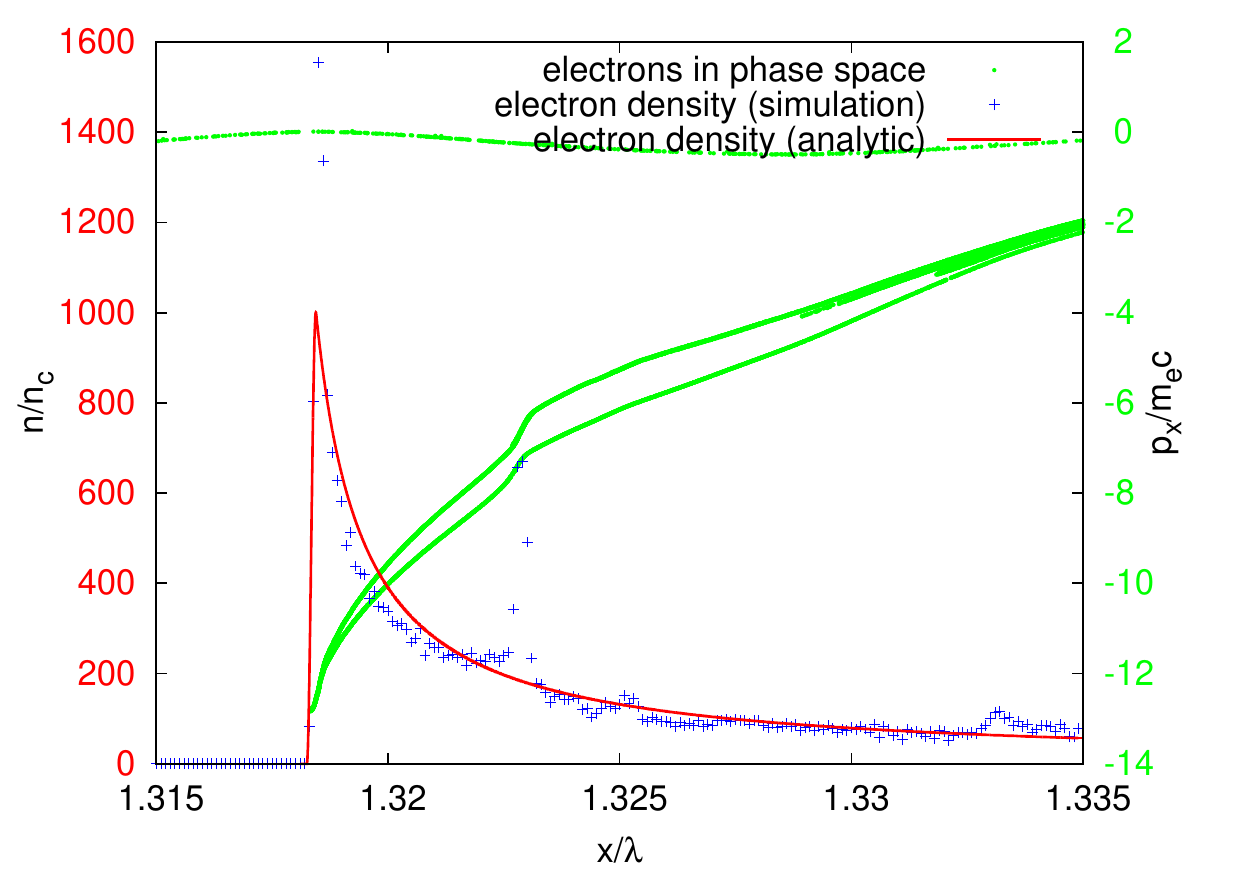}
		\caption{\small Electron density taken from simulation (blue) and calculated analytically via (\ref{profile_exp}) (red), as well as electrons in $x$-$p_x$-plane (green),  
		with same simulation parameters compared to Fig. \ref{xpx2}, taken at $t=5.7\lambda/c$. 
		$x_\text{min}-x_0=9\cdot10^{-4}\lambda$; $x_\text{max}-x_0=0.02\lambda$; $a=1\cdot10^{-4}$ (simulation frame). }
		\label{fit_dens_exp}
	\end{figure}

	Now as in the previous case we are going to compare the calculated analytical function with the simulated density profile (Fig. \ref{fit_dens_exp}).
	Again we obtain a good agreement and are able to describe the density spike quite well. 

\begin{figure}[htb]
		\centering
		\includegraphics[width=7cm]{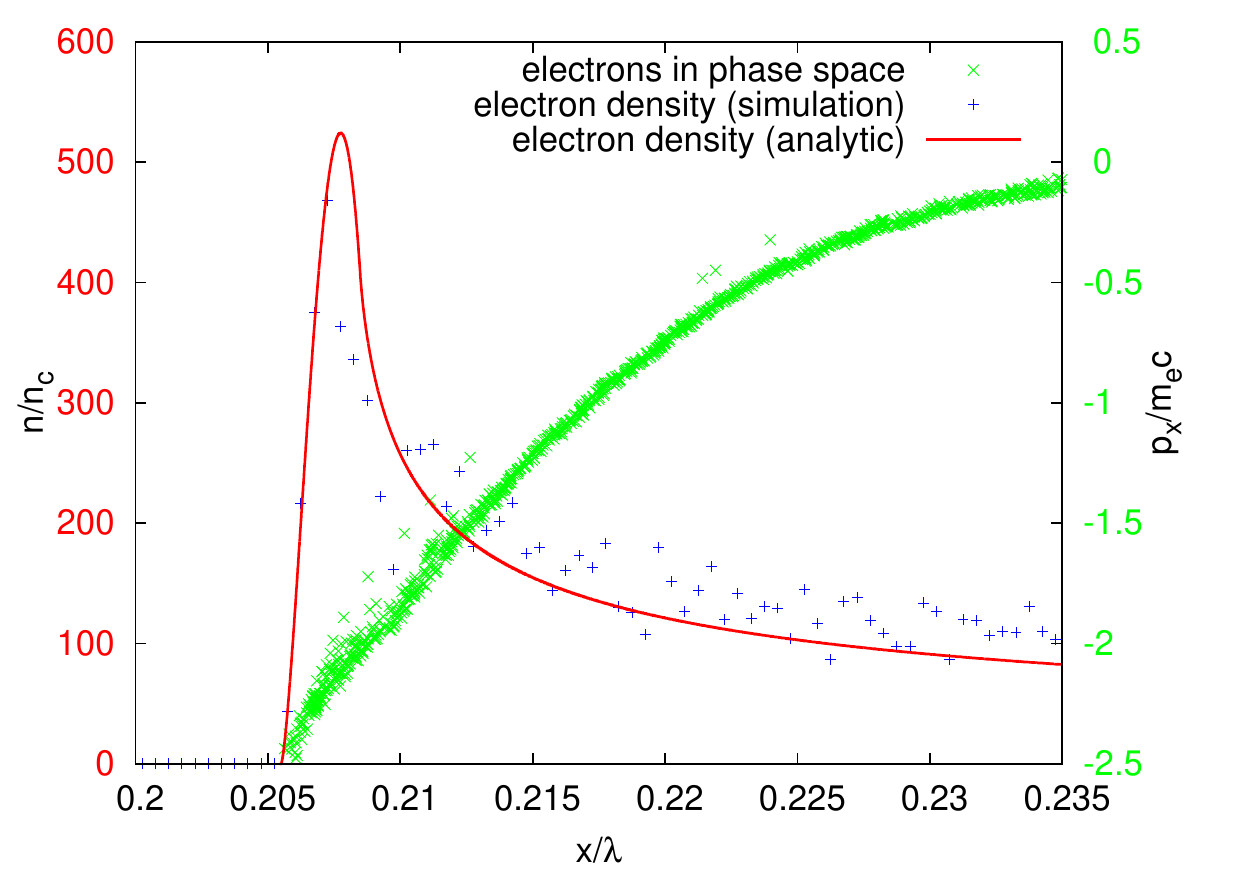}
		\caption{\small Electron density taken from simulation (blue) and calculated analytically via (\ref{profile_sqrt}) (red), as well as electrons in $x$-$p_x$-plane (green).  
		Simulation parameters: initial plasma density $n_0 = 100 n_c$; $\sigma=0.066\lambda$ (laboratory frame), Pulse with dimensionless amplitude $a_0 = 10$
		and p-polarized oblique incidence at $60^\circ$ angle has the  wave length $\lambda = 820 $nm.}
		\label{fit_dens_sqrt}
	\end{figure}

	Before we go further to the next chapter we analyze the intermediate case $\dot x_0(t) \lesssim c $, which is important for further application. 
	In this case the electron phase space distribution looks like its shown in Fig. \ref{fit_dens_sqrt} and can't be approximated well either with parabolic nor with exponential function. 
	Nevertheless we find out that the density profile of the spike can still be well approximated with equation (\ref{profile_sqrt}) (Fig. \ref{fit_dens_sqrt}), 
	so we classify the cases with intermediate velocities as parabolic.
	
	In the following chapter we are going to analyze corresponding simulation results more extensively.  
	We will use the descriptions of the electron layer density profile derived here in order to calculate an expression for the spectra of the reflected waves in different cases. 
	
\section{Electron density evolution and HHG emission}	
	We are interested in high frequency spectrum of the reflected pulse mostly determined by the behavior of the ARP when it moves away from plasma with maximal velocity. 
	This moment corresponds to a stationary phase point (SPP) (see  \cite{BP2, BP}). The ARP gamma factor exhibits a sharp spike in this time, it is called $\gamma$-spike \cite{BGP}. 
	One distinguish different orders of $\gamma$-spikes depending on behavior of the  transverse current in vicinity of SPP, which can by approximated with
	\begin{equation}
		j_\bot(t,x)\approx (-\alpha_0 t)^nf(x-x_0(t)).
		\label{j1}
	\end{equation}	
	We assume that the transverse current density doesn't change its shape $f$ during the time. The number $n$ denotes order of the given $\gamma$-spike. 
	The reflected radiation is determined by the transverse current distribution via
	\begin{equation}
		E_r(t)=\pi\int j_\bot(t-x,x)dx,
		\label{er2}
	\end{equation}
	so we are able to derive the expression for the spectrum of the reflected pulse in line with \cite{BP2, BP} and obtain: 
	\begin{align}
		I(\omega)&=4\pi^4\alpha_0^2(\alpha_1\omega)^{-\frac{2n+2}{2n+1}}\left(\frac{d^n}{d\xi^n} Ai_n(\xi_n)\right)^2 |f(\omega)|^2,
		\label{Ai}
		\\
		\nonumber
		\xi_n&=\alpha_1^{-\frac{1}{2n+1}}\delta\omega^\frac{2n}{2n+1}, \quad Ai_n=\frac{1}{2\pi}\int e^{i\left(xt+\frac{t^{2n+1}}{2n+1}\right)}dt,
		\\
		\nonumber
		\alpha_1&=\frac{a_0^2}{2\upsilon n_m^2},\quad\delta=1-\upsilon,
	\end{align}
	where $\upsilon$ is the maximal velocity of the moving electron layer in SPP and $n_m$ the maximal density assumed to be constant in time. 
	In (\ref{er2}) we use the normalized PIC units, see \cite{PIC} for more detail.
	To give  the expression for the shape function we use the results from the previous section and write
	\begin{equation}
		f(x)=\frac{n_a(x)}{n_a(x_\text{m})}e^{-\frac{x^2}{\tilde{\tilde{\sigma}}^2}},\quad n_a(x_\text{m})=n_m.
		\label{f}
	\end{equation}
	We multiply the density profile with wider Gaussian function since $n_a$ decays too slowly ($\propto1/x$ whip or $\propto1/\sqrt{x}$ parabolic) for positive $x$ 
	and after certain $x$-value doesn't coincide with given density. 
	
	Further we consider two different examples were we apply (\ref{Ai}) to calculate the spectrum of the single reflected pulse $E_r^\text{pls}(t)$, 
	that is   filtered out by the Gaussian function
	\begin{equation}
		E_r^\text{pls}(t)=E_r(t)e^{(t-t_\text{max})^2/\tilde{\sigma}^2}, 
	\end{equation}	
	where $t_\text{max}$ corresponds to the maximal wave amplitude and $\tilde{\sigma}=0.2\lambda/c$.

\begin{figure}[htb]
		\centering
		\includegraphics[width=7cm]{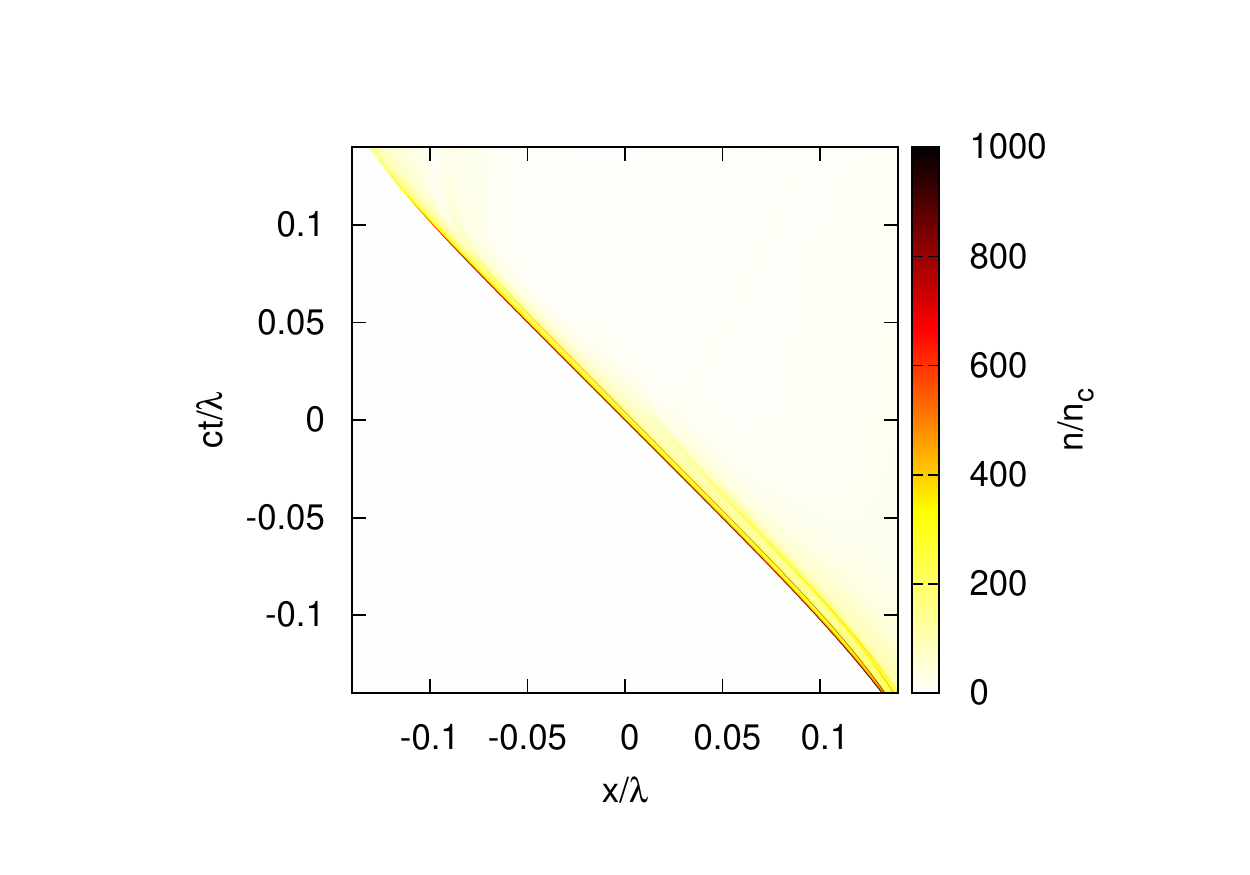}
		\caption{\small The electron density distribution of the radiating nanobunch in space time domain.
		Simulation parameters  are the same compared to Fig. \ref{xpx2}. 
		The SPP (here (0,0)) corresponds to $t=5.7\lambda/c$  as in Fig. \ref{fit_dens_exp}.}
		\label{dens1_2D}
	\end{figure}

\begin{figure}[t]
		\centering
		\includegraphics[width=8.6cm]{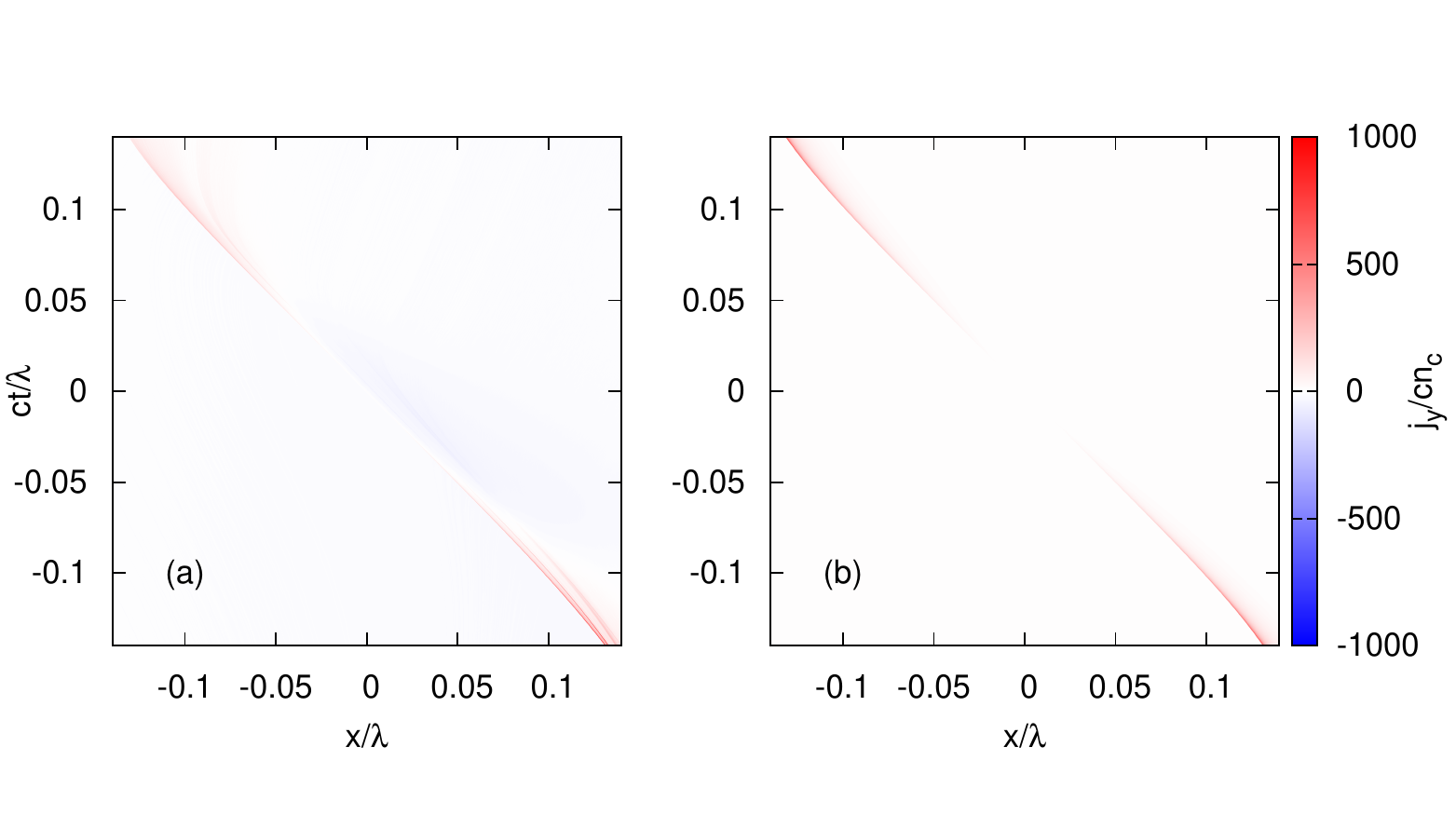}
		\caption{\small Transverse current density from the simulation near the SPP (0,0) (a) and calculated analytically via. (\ref{j1}), $n=2$ (b).
		Simulation parameters  are the same compared to Fig. \ref{xpx2}. 
		The parameters used by the analytical calculation for $x_0(t)$: $\alpha_0=4\cdot10^4$, $n_\text{m}=1000$ and $\gamma=15$,
		for shape: $a=1\cdot10^{-4}\lambda$, $x_\text{min}=9\cdot10^{-4}\lambda$ and $\tilde{\tilde\sigma}=0.02\lambda$. 
		The velocity $\upsilon$  is derived from the given gamma factor. The Fourier transform of the shape function $f(\omega)$ is calculated numerically using FFT.}
		\label{jy_2D}
	\end{figure}

\begin{figure}[htb]
		\centering
		\includegraphics[width=8.6cm]{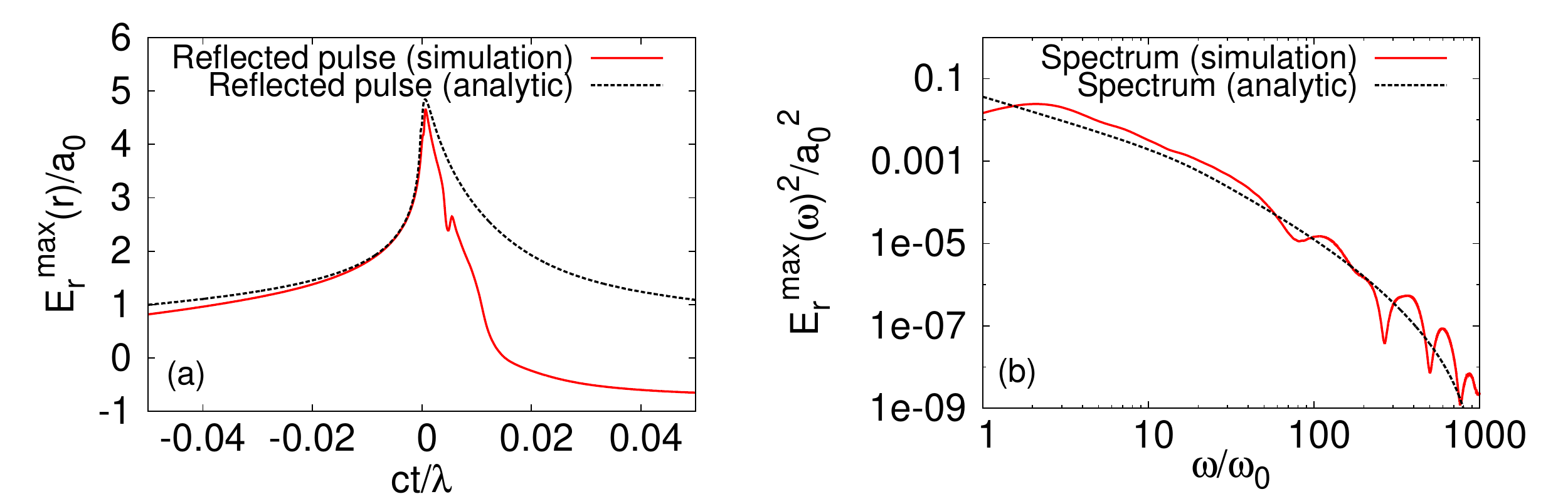}
		\caption{\small Reflected radiation obtained from the simulation ((a) red) and from analytical current distribution ((a) black), as well as the corresponding spectra in (b). 
		The spectrum from the simulation is taken directly from the radiated pulse via FFT, while the other one is obtained using the equation (\ref{Ai}). }
		\label{airy}
	\end{figure}

	At first we investigate the example of the whip case ($\dot x_0(t)\rightarrow c$) from the previous section illustrated in  
	Fig. \ref{xpx2} and Fig. \ref{fit_dens_exp}  more extensively. 
	The electron nanobunch which radiates strong attosecond pulse can be clearly recognized from Fig. \ref{dens1_2D} and \ref{jy_2D} (a). 
	For convenience we chose the coordinates in the way that the SPP is in the point (0,0). 
	In Fig. \ref{airy} (b) the spectrum calculated using (\ref{Ai}) is compared with the spectrum calculated von original reflected Pulse via FFT.  
	Obviously, the description works well almost until 1000-th harmonic. 
	In Fig. \ref{airy} (a) the corresponding pulses are compared. 
	The both graphs behave in the similar manner.

\begin{figure}[htb]
		\centering
		\includegraphics[width=7cm]{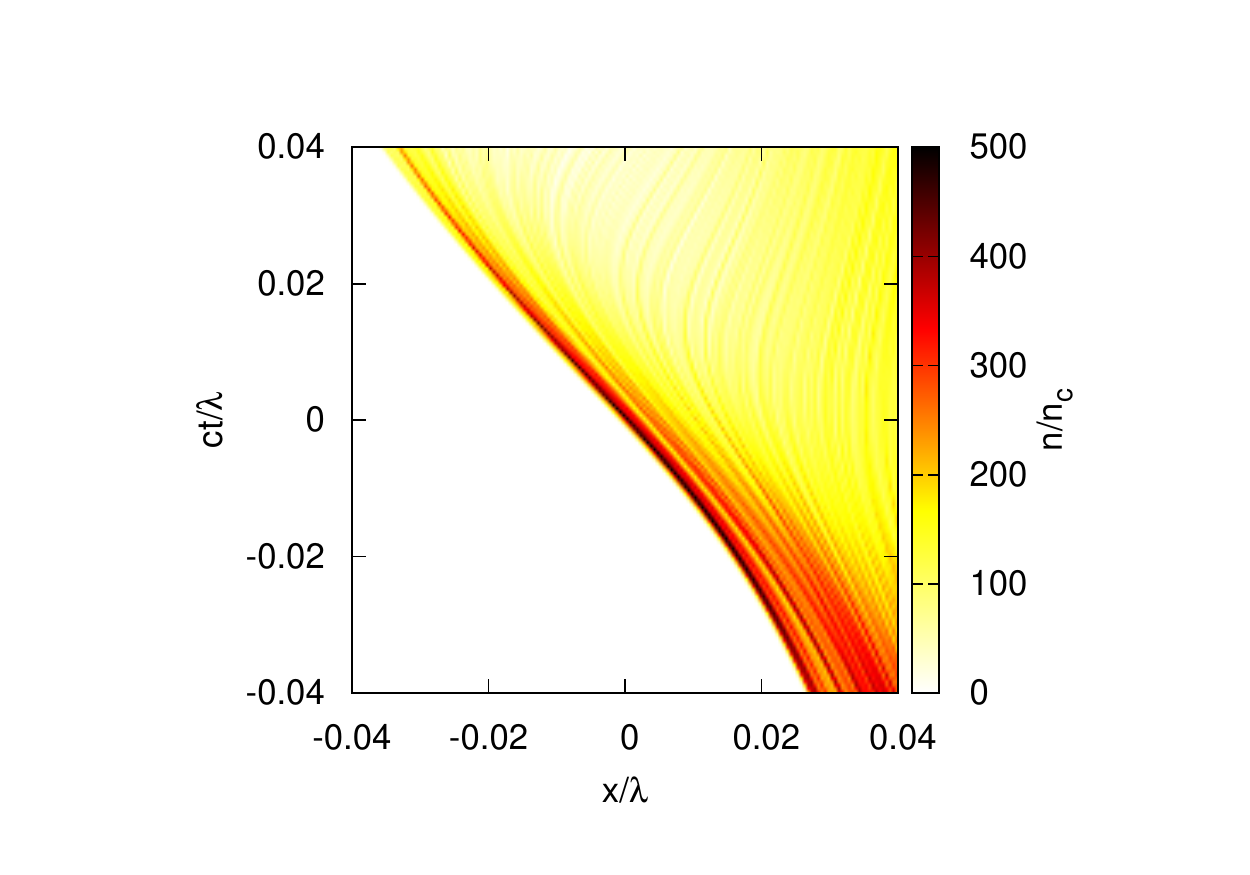}
		\caption{\small
		 The electron density distribution of the radiating nanobunch the  in space time domain.
		Simulation parameters  are the same compared to Fig. \ref{fit_dens_sqrt}. 
		The density profile in SPP (here (0,0)) is shown  in Fig. \ref{fit_dens_sqrt}.}
		\label{dens1_gamma_2D}
	\end{figure}

\begin{figure}[htb]
		\centering
		\includegraphics[width=8.6cm]{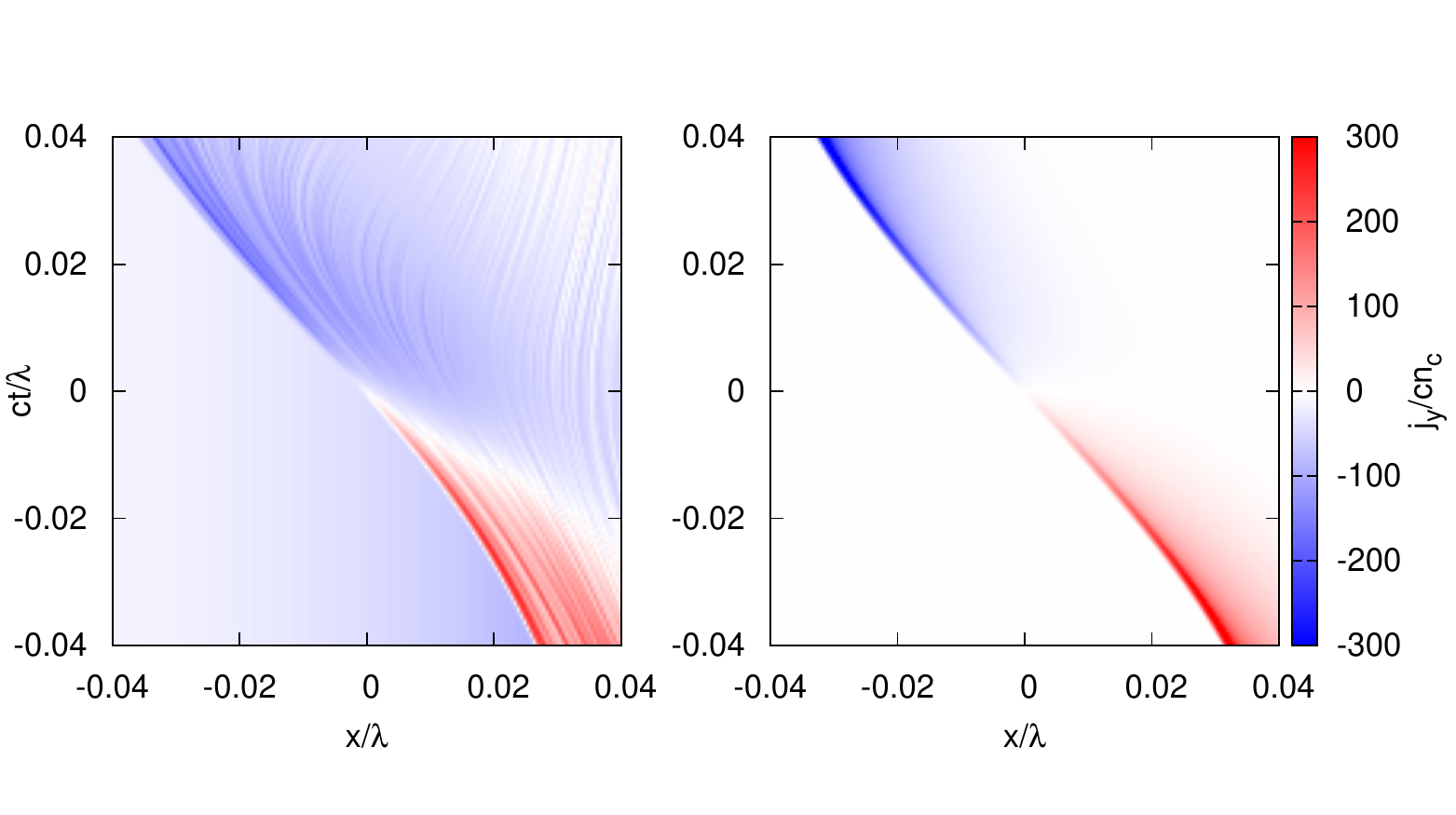}
		\caption{\small Transverse current density from the simulation  near the SPP (a) and calculated analytically via. (\ref{j1}), $n=1$ (b).
		Simulation parameters  are the same compared to Fig. \ref{fit_dens_sqrt}. 
		The parameters used by the analytical calculation for $x_0(t)$: $\alpha_0=1\cdot10^4$, $n_\text{m}=500$ and $\gamma=2.5$, 
		for shape: $a=1\cdot10^{-3}\lambda$ and $\tilde{\tilde\sigma}=0.02\lambda$. The velocity $\upsilon$  is derived from the given gamma factor.}
		\label{jy2_2D}
	\end{figure}

\begin{figure}[ht]
		\centering
		\includegraphics[width=8.6cm]{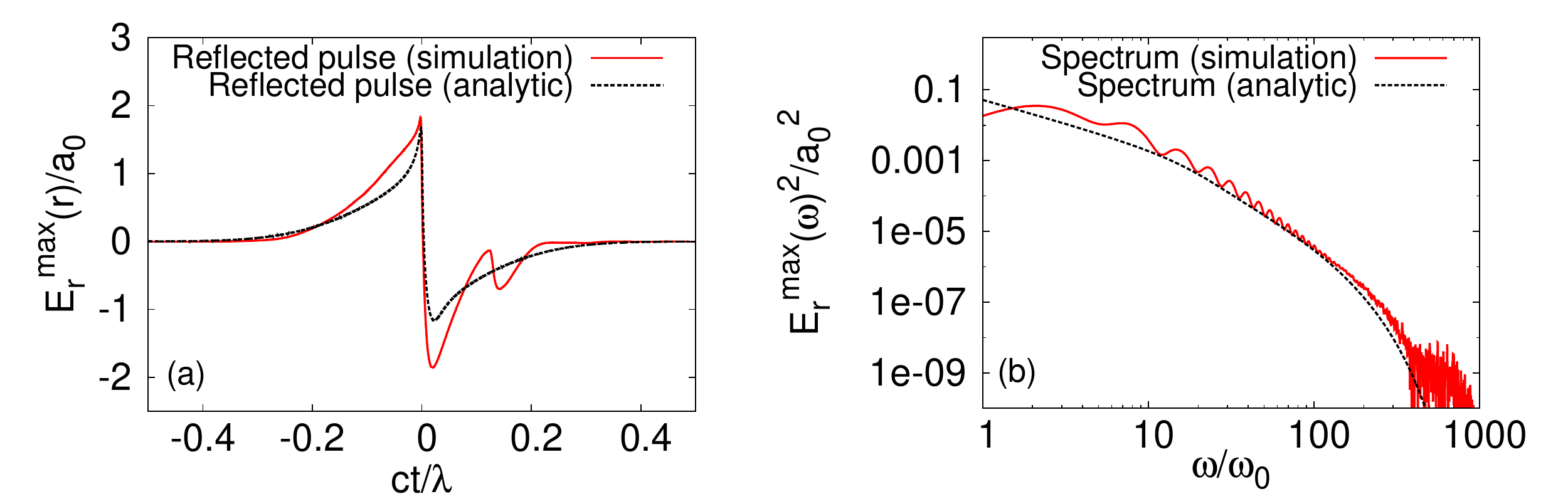}
		\caption{\small Reflected radiation obtained from the simulation ((a) red) and from analytical current distribution ((a) black), as well as the corresponding spectra in (b). }
		\label{airy1}
	\end{figure}

	Going along the same lines we analyze now the intermediate case $\dot x_0(t) \lesssim c$ shown in Fig. \ref{fit_dens_sqrt}. 
	Es we said before we attribute this case to the parabolic case.  
	The corresponding pictures illustrating this case are Fig. \ref{dens1_gamma_2D}, \ref{jy2_2D} and \ref{airy1}.   
	Here the velocity in SPP significantly deviates from the speed of light and approximately equals to $0.91c$.  
	For that reason the electron phase space distribution doesn't become ``whip-like'' (Fig. \ref{fit_dens_sqrt}). 
	Although there is no ultrarelativistic regime here, we still may apply the same analysis assuming the absolute velocity of the electrons to be approximately  constant  close to SPP. 
	Again we obtain good agreement for the spectrum behavior. 	

\begin{figure}[htb]
		\centering
		\includegraphics[width=7cm]{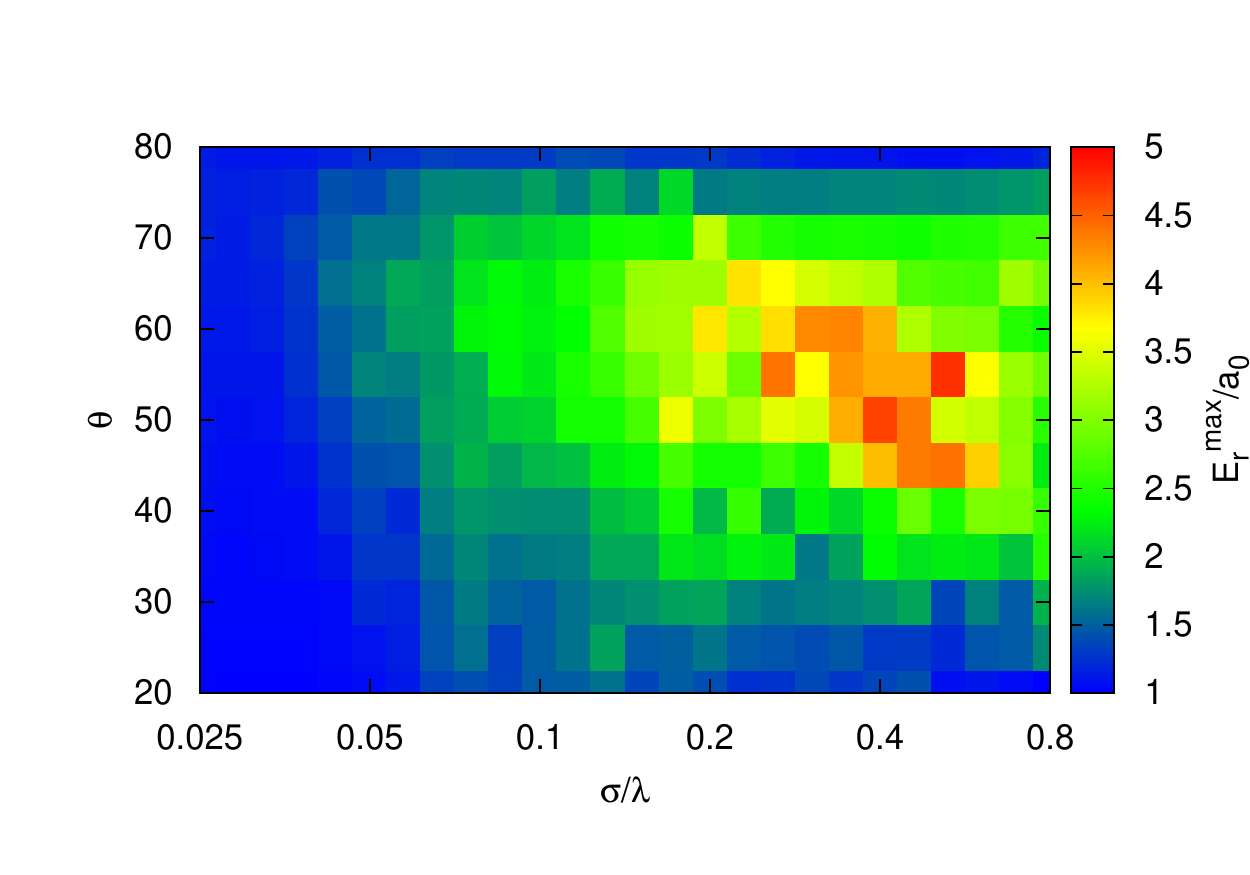}
		\caption{\small Each point in these pictures corresponds to the maximal amplitude obtained from the reflected radiation taken from corresponding simulation. 
		We have different angles of incidence along $y$-axis and different steepness of the density gradient along $x$-axis, where $\sigma$ is taken from (\ref{rump})
		and $n_0=100n_c$ (laboratory frame). }
		\label{params}
	\end{figure}

	We see that  in the first example (whip case) we have the second order gamma spike, whether in the second example (parabolic case) the first order gamma spike is obtained. 
	In order to find the parameter ranges which correspond to whip or parabolic case we perform a number of simulations. 
	Using moderate intensity of incident wave ($a_0=10$) we vary the steepness of the exponential density gradient as well as incident angle. 
	For each parameter set we consider the reflected radiation. 
	In Fig. \ref{params} we visualized the maximal amplitude of reflected wave for each parameter set respectively.  	
	Consider the incident angle between 45 and 60, since by this angles the most interesting things happen. 
	Of corse we notice the sharp increase of the reflected wave amplitude in the area around $\sigma=0.4\lambda$. 
	We call this area high  amplitude parameter set (HAPS).   
	In this area we mostly obtain the second order $\gamma$-spikes  and the current doesn't change its sign in SPPs like in Fig. \ref{jy_2D} 
	Furthermore our study shows that the maximal longitudinal velocity of the certain boundary electron layer increase monotonously with $\sigma$ until HAPS, 
	where it almost reaches $c$. For $\sigma<0.05\lambda$ the boundary oscillates to slowly so that any short pulses are generated. 
	Roughly between $0.05\lambda$ and $0.1\lambda$ we obtain the reflected radiation similar to Fig. \ref{airy1}. 
	We call this area moderate amplitude parameter set (MAPS). Here we have only first order $\gamma$-spikes and the current changes sign in SPPs (Fig. \ref{jy2_2D}). 
	Thus the reflected spectrum in MAPS can be approximated with  equation (\ref{profile_sqrt}) (parabolic case) 
	and the area of HAPS corresponds than to the exponential case (equation (\ref{profile_exp})). 
	In the area between MAPS and HAPS the interaction is too complicated to be attributed to any model.

\section{conclusion}
	We could obtain two different   analytical expressions of electron density profile describing the density spikes in two different cases.  
	%Further we used this expression to calculate back radiation of moving electron  layer and compared our result with one from \cite{GKMS}. 
	Further we presented some simulation results of HHG, where we could obtain the  amplitude increasing  in the reflected pulse by the factor of five
	without using extremely intense incident wave. 
	This was possible after we found optimal parameters for density gradient combined with optimal incident angle.  
	Moreover with some simple assumptions we were able to describe the distribution of transverse current in vicinity of SPP analytically in both cases. 
	Obtained expressions together with the expressions for electron density  gave us the possibility to calculate the spectra that fits the original spectra of back radiated pulse quite good.  

	Our work basically presents the idea of description of the plasma density considering the electron  phase space distribution, 
	but this theory has a potential to grow and to be developed further.   In this work we introduced just two application examples of our theory. 
	On the other hand it can become a strong tool in laser plasma analysis in general.

This work has been supported by DFG TR18 and EU FP7 Eucard-2 projects.

\bibliography{apssamp}

\end{document}